\documentclass[epj]{svjour}
  \usepackage{graphicx}
  \usepackage{amsmath}
 \usepackage{amsfonts}
 \usepackage{amssymb}
\usepackage{pstricks}
\usepackage{psfrag}
\usepackage{epsfig}
\usepackage[ansinew]{inputenc}
  
\renewcommand{\vec}[1]{\mathbf{#1}}

\begin{document}
\title{Traveling spatially periodic forcing of phase separation}
\author{Vanessa Weith, Alexei Krekhov
\and Walter Zimmermann
}                     
\authorrunning{V. Weith, A. Krekhov, W. Zimmermann}
\institute{Theoretische Physik I, Universit\"at Bayreuth, D-95440 Bayreuth, 
            Germany}
\date{Dated: \today}
%

\abstract{We present a theoretical analysis of phase separation in the
presence of a spatially periodic forcing of wavenumber $q$ traveling
with a velocity $v$. By an analytical and numerical study of
a suitably generalized 2d-Cahn-Hilliard model we find
as a function of the forcing amplitude and the 
velocity three different regimes of phase separation.
For a sufficiently large forcing amplitude a spatially
periodic phase separation of the forcing wavenumber
takes place, which is dragged by the forcing 
with some phase delay. These locked solutions are
only stable in a subrange of their existence and beyond
their existence range the solutions are
dragged irregularly during the initial transient
period and otherwise rather regular. In the range
of unstable locked solutions a coarsening dynamics
similar to the unforced case takes place. For
small and large values of the forcing
wavenumber analytical approximations of the nonlinear solutions
as well as for the range of existence and stability have been
derived.
}
\maketitle

\section{Introduction}\label{sec: intro}
When a binary mixture at a critical composition is quenched
from a homogeneous and usually high temperature phase sufficiently below the coexistence
curve then inhomogeneous concentration variations of a characteristic average domain size occur which are known as
spinodal decomposition (SD) \cite{Gunton:83.1,Bray:94.1}.
With the progress of time a coarsening process takes place, i.e. the average domain size
increases continuously. The dynamics of phase separation
under homogeneous conditions has been intensively studied for numerous
systems during the recent decades \cite{Gunton:83.1,Bray:94.1}.
In this work we explore the effects of a traveling spatially periodic forcing on phase separation.

For several kinetic mechanisms associated
with first order phase transitions, such as nucleation and growth,
thermal activation energy is required, but not for SD. The
order parameter, that describes the phase separation in binary mixtures, is 
the local composition (concentration) and it obeys a local conservation
law. Hence in phase separating systems the order parameter is a
conserved one in contrast to the large number of 
pattern forming systems, where one has to deal with
non-conserved order parameters \cite{CrossHo}.
The phase separation is limited by diffusion-like processes
and therefore in its late stage, where small inhomogeneities have
developed into macroscopic domains, the decomposition dynamics 
becomes rather slow.

Phase separation is also employed in producing technologically important functional
elements in e.g. lithographic processes, biosensors and semiconductor devices \cite{Walheim:1999.1,Boeltau:1998.1,Voeroes:2005.1,Sirringhaus:2005.1}.
Therefore, an understanding of the mechanisms of the formation of
regular structures is highly important for technological applications
and even the control of regular patterns is desired.

In a large number of investigations the effects of external fields
on phase separation have been examined, 
 such as for instance the effects of shear flow \cite{Berthier:2001.1}, time-dependent temperature variation
 \cite{Tanaka:1995.1} or
local \cite{Krekhov:2005.1} as well as spatially periodic heating
\cite{Krekhov:2004.1}. In the presence of spatially periodic forcing 
coarsening can be interrupted above a critical
modulation amplitude and locked solutions can be found \cite{Krekhov:2004.1}.

Especially for pattern forming systems with non-conserved order parameters 
external forcing has been recognized as a powerful method for investigations
of the response behavior of nonlinear patterns and of the inherently
nonlinear mechanisms of selforganization. The effect of spatially
periodic forcing on pattern formation has been investigated rather early in the 
context of thermal forcing \cite{Kelly:78.1} and electroconvection
in nematic liquid crystals \cite{Lowe:83.1,Lowe:86.1}. Further on, the
response of stationary patterns with respect to static spatially
periodic forcing in one or two spatial dimensions has 
been explored with respect to modulated structures \cite{Lowe:83.1,Lowe:86.1,Coullet:86.2,Coullet:87.1,Coullet:89.1,Zimmermann:93.3,Zimmermann:96.1,Kai:1999.1,Ribotta:2003.1} or with respect to induced time-dependent phenomena \cite{Rehberg:91.2,Zimmermann:96.2,Zimmermann:96.3}. Separately also
the effects of temporal forcing have been investigated \cite{Riecke:88.1,Walgraef:88.1,Rehberg:88.1,Swinney:2000.1,Swinney:2000.2}.
Recently a number of new phenomena have been found and explored
by a combination
of spatial and temporal forcing in model systems \cite{Krekhov:2008.1,Zimmermann:2002.2,Sagues:2003.1,Kramer:2004.1,Schuler:2004.1,Ruediger:2007.1,Ruediger:2007.2}.
Here we present the first investigation of the effects of pulled
spatially periodic external forcing on a pattern forming model
system with a conserved order parameter.

The paper is organized as follows.
In Sec. 2 we present the model equation for the conserved order parameter field. In
section 3 the one-dimensional periodic solutions and their existence range are
described and their linear stability analysis in one and two dimensions is presented 
in section 4. The results of numerical simulations of the dynamics of phase separation
under traveling spatially periodic forcing in 1D and 2D are given in Sec. 5.
Some concluding remarks are added in Sec. 6. In the Appendix we describe
the condition for the existence of spatially periodic solutions and analytical approximations
for the solutions in the limit of small and large values of the forcing wavenumber.

\section{Model}\label{sec: numeric}
A model for
phase separation in various systems is the Cahn-Hilliard equation for a conserved order parameter field
$\psi(\vec{r},t)$  representing the local concentration
of one of the components of a binary mixture.
Here we focus on the effects of traveling spatially periodic forcing on
phase separation as described by the modified Cahn-Hilliard
equation
\begin{equation}
 \partial_t\psi(\vec{r},t)=\nabla^2\lbrace-\varepsilon\psi+\psi^3-
\nabla^2\psi+a\cos[q(x-vt)]\rbrace.\label{basicequation1}
\end{equation}
The forcing term $a\cos[q(x-vt)]$ is caused by an interplay
between a traveling temperature modulation and thermal diffusion (Soret effect).
Its derivation has been described in detail for $v=0$ in Ref.~\cite{Krekhov:2004.1}.
Such a traveling spatially periodic temperature modulation could be created for instance in
optical grating experiments \cite{Wiegand:2002.1,Wittko:2003.1} with a light intensity
$I(x,t)\sim \cos[q(x-vt)]$. The control parameter $\varepsilon$ in Eq.~(\ref{basicequation1})
corresponds to a dimensionless distance to the critical temperature of the binary mixture. 
A transformation of Eq.~(\ref{basicequation1}) in the frame comoving with the
traveling forcing $x\rightarrow x-vt$ gives
\begin{equation}
 \partial_t\psi(\vec{r},t)=\nabla^2[-\varepsilon\psi+\psi^3-
\nabla^2\psi+a\cos(qx)]+v\,\partial_x\psi.\label{basicequation2}
\end{equation}

\section{Periodic solutions and bifurcation diagrams in 1D}\label{sec: results}
During the initial stage of an unforced phase separation, i.e. $a=0$ and $v=0$, small 
perturbations of the homogeneous solution $\psi=0$ exhibit an exponential 
growth for $\varepsilon>0$ and the growth rate $\sigma=k^2(\varepsilon-k^2)$
shows a maximum at a distinguished wavenumber $k_m=\sqrt{\varepsilon/2}$.
Beyond the regime of exponential growth
one observes the well known
coarsening dynamics of phase separation. 

If phase separation is forced by a spatially periodic temperature modulation the
coarsening dynamics is interrupted beyond a critical
value of the forcing amplitude $a$ and it is locked 
to the periodicity of the external forcing \cite{Krekhov:2004.1}. However, if this forcing is
pulled by a velocity $v\not =0$, the spatially periodic solutions
do exist only in a certain range of $v$ depending on $a$. The conditions for the existence
of spatially periodic solutions are formulated in Appendix \ref{appA1}
and they are explicitely calculated for various parameters in this section.

The bifurcation diagrams of spatially periodic solutions of Eq.~(\ref{basicequation2}) 
are described for $v=0$ in subsection \ref{statnonprop} as well as 
for $v\neq0$ in \ref{statprop} and the numerical method for
calculating the solutions is explained in subsection \ref{numnonlin}.

\subsection{Numerical solution} \label{numnonlin}
The stationary $2\pi/q$-periodic solutions of Eq.~(\ref{basicequation2}) can be represented by
a truncated Fourier expansion
\begin{equation}
\label{Fourieransatz}
 \psi_s(x)=\sum\limits_{k=-N}^N C_k e^{iqxk}
\end{equation}
with time-independent Fourier coefficients $C_{-k}=C_k^*$.
 Using this ansatz one obtains after projection 
of Eq.~(\ref{basicequation2}) onto the 
$j$-th Fourier mode ($\int_0^{2\pi/q}e^{-iqxj}dx$), the coupled system of nonlinear inhomogeneous equations for the corresponding
expansion coefficients $C_j$
\begin{eqnarray}
 &&[ \varepsilon (jq)^2  -(jq)^4 +i v jq]C_j  - \frac{aq^2}{2}(\delta_{1,j}+\delta_{-1,j}) 
 \nonumber \\
\nonumber \\
&& \qquad -(jq)^2\sum\limits_{k,l} C_k C_l C_{j-k-l} =0\,,\,\,\,\,\,\,j=-N..N,\label{statlsg}
\end{eqnarray}
which has been solved
by Newton's method. $N$ was adjusted in order to keep a relative error smaller than $10^{-6}$
(e.g. $N=81$ for $q=0.5$).

\subsection{Periodic solutions for $v=0$} \label{statnonprop}
\begin{figure}[ht]
  \begin{center}
\includegraphics[width=0.95\columnwidth]{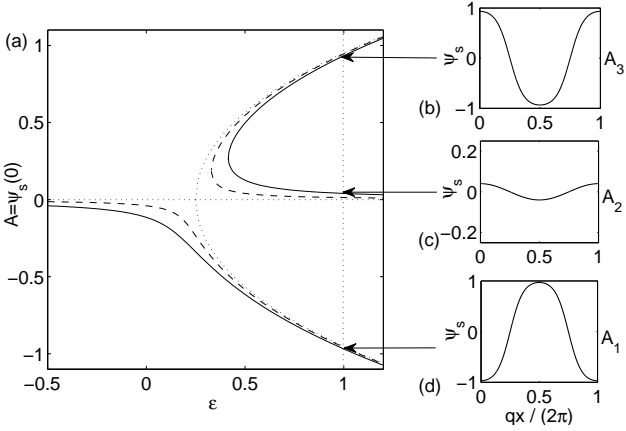}
  \end{center}
  \caption{Part (a) shows the bifurcation diagram for spatially periodic
solutions at a forcing wavenumber  $q=0.5$ and modulation 
amplitudes $a=0$ (dotted), $a=0.01$ (dashed) and $a=0.03$ (solid). 
Parts (b), (c) and (d) show the three solutions $\psi_s(x)$ over one period 
corresponding to the different branches of the bifurcation 
diagram for $a=0.03$ and $\varepsilon=1$.}
\label{bif1}
\end{figure}

For positive values of $\varepsilon$ spatially periodic 
solutions of Eq.~(\ref{basicequation2}) may
be found with and without forcing. 
However, in the unforced case ($a=0$)
Eq.~(\ref{basicequation2}) 
has a $\pm \psi$-symmetry and therefore one has a pitchfork bifurcation from
the trivial solution $\psi=0$ to finite amplitude periodic solutions as 
indicated by the dotted line in Fig.~\ref{bif1}. For $a=0$ periodic solutions are unstable
for any wavenumber $q$ against coarsening with the fastest growing mode as period doubling \cite{Langer:71.1,Krekhov:2004.1}.
 
For finite values of the modulation amplitude $a$
the $\pm \psi$-symmetry is broken as indicated in Fig.~\ref{bif1}
for $a=0.01$ by the dashed line and for $a=0.03$ by the solid line.
While we have in the unmodulated case a trivial solution
$\psi=0$ and
two finite solutions with identical amplitudes but of opposite sign,
one finds in the forced case three periodic solutions, 
$A_1$, $A_2$ and $A_3$,  of different amplitude 
as shown in Fig.~\ref{bif1} (b)-(d). The $A_3$- and $A_2$-solutions are in phase
with the external modulation and the preferred $A_1$-solution is shifted by half 
a period.

\begin{figure}[ht]
  \begin{center}
\includegraphics[width=0.98\columnwidth]{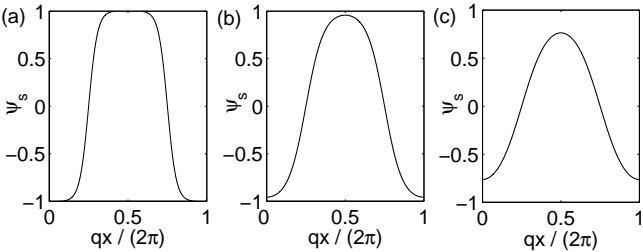}
  \end{center}
  \caption{$A_1$-solution for different 
values of the modulation wavenumber: $q=0.25$ (a), $q=0.5$ (b)
and $q=0.75$ (c). The other parameters are $\varepsilon=1$, $v=0$ and $a=0.01$.}
\label{bif2}
\end{figure}

The amplitudes of the $A_1$- and the $A_3$-solution increase with increasing values of the control parameter $\varepsilon$, 
but  they also depend on $q$ and $a$. 
For a given modulation amplitude $a$ the amplitude of $A_1$ decreases 
with increasing values of the
modulation wavenumber $q$ as indicated in 
Fig.~\ref{bif2}. This trend holds also for the
$A_3$-solution but the amplitude of the $A_2$-solution increases with $q$.

Keeping $q$ fixed and increasing the modulation amplitude $a$
the amplitude of the $A_1$-solution increases as can be seen in Fig.~\ref{bif1}.
This also holds for the $A_2$-solution.
In contrast to this trend, the amplitude of the $A_3$-solution
decreases with increasing values of $a$ (see dashed and solid lines in Fig.~\ref{bif1}).
In case of small $q$ (long wavelength modulation) the solutions $A_1$ and $A_3$ become steplike as
shown in Fig.~\ref{bif2} (a) for $A_1$. This steplike shape also occurs, if 
the control parameter $\varepsilon$ becomes rather large compared to 
the modulation amplitude. The shape of the $A_2$-solution does not change when $a$ or $\varepsilon$
are varied.

\subsection{Periodic solutions for $v\not =0$}\label{statprop}
\begin{figure}[ht]
  \begin{center}
\includegraphics[width=0.98\columnwidth]{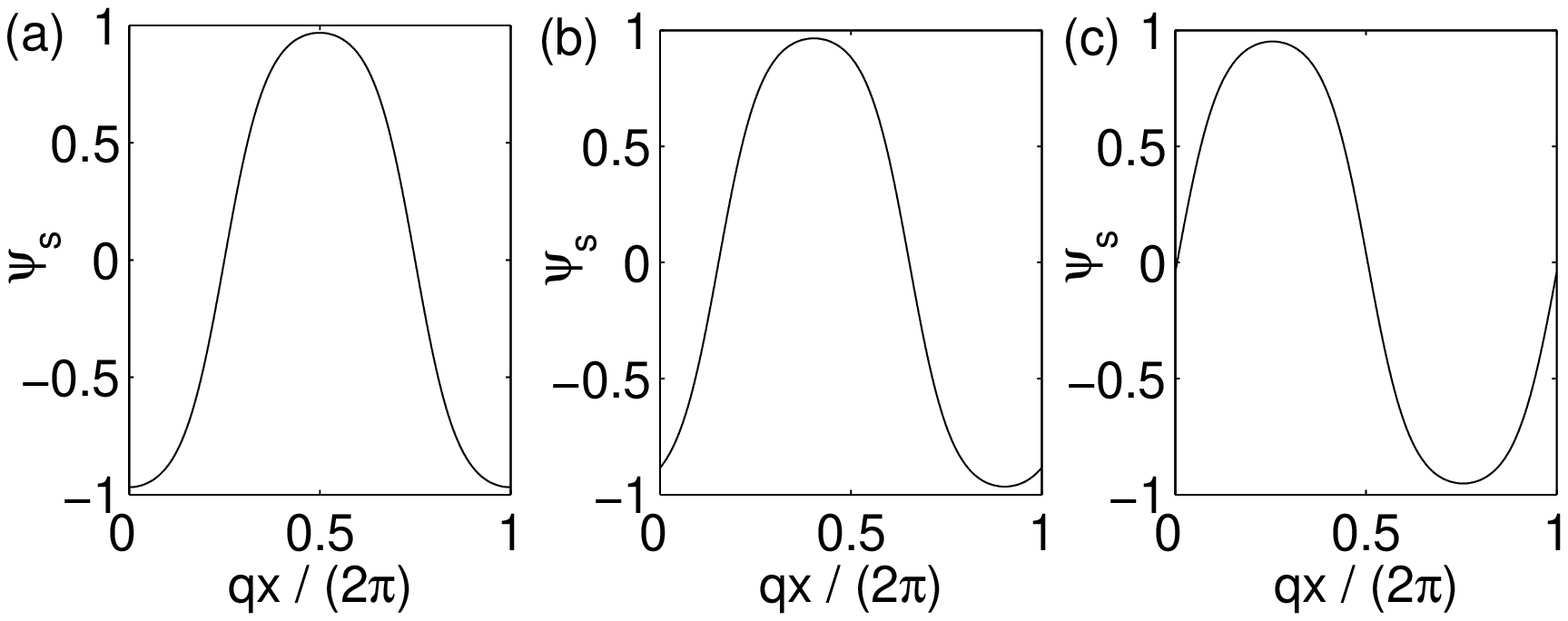}\\
\includegraphics[width=0.98\columnwidth]{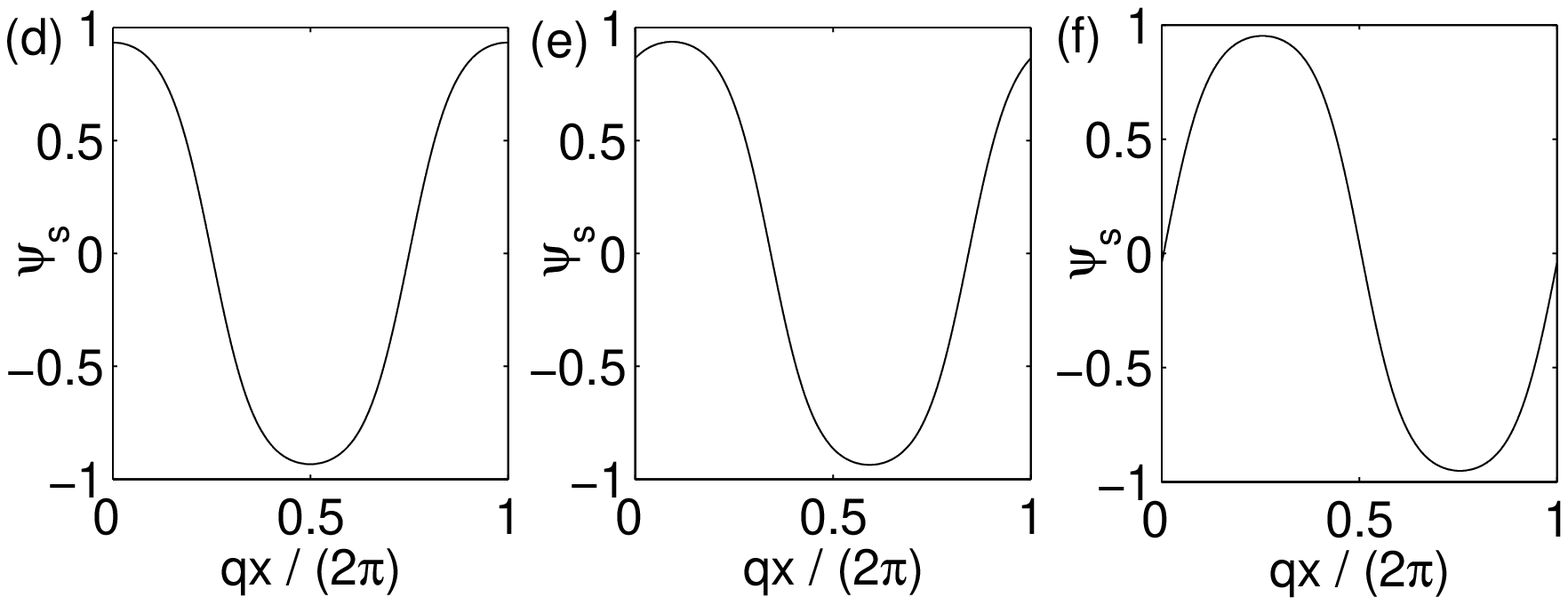}
  \end{center}
  \caption{$A_1$-solution (top) and $A_3$-solution (bottom) for 
the velocity $v=0$ (a,d), $v=0.008$ (b,e)
and $v=0.0142$ (c,f). The other parameters are $q=0.5$, $a=0.03$ and $\varepsilon=1$.
}
\label{bifv1}
\end{figure}

\begin{figure}[ht]
  \begin{center}
\includegraphics[width=0.95\columnwidth]{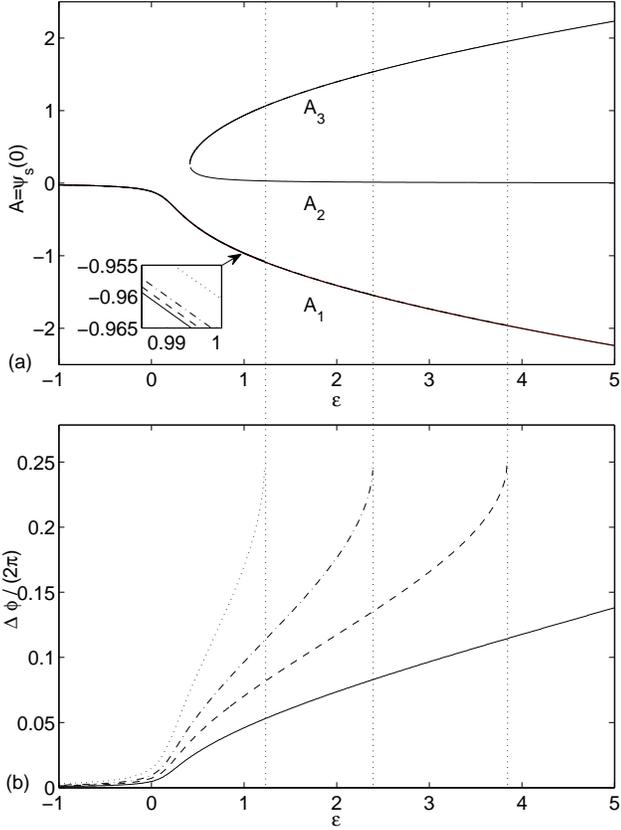}
  \end{center}
  \caption{The bifurcation diagram for $q=0.5$, $a=0.03$ and finite values of the velocity $v$ is shown in (a) and in (b) the
phase shift $\Delta \phi/(2\pi)$ of the $A_1$-solution is presented for different velocities: $v=0.004$ (solid), $v=0.006$ (dashed),
$v=0.008$ (dashed-dotted) and $v=0.0142$ (dotted).}
\label{bifv2}
\end{figure}

The nonlinear solutions and the
bifurcation diagram as given in Fig.~\ref{bif1}  are only
slightly changed by a finite traveling velocity. 
However, with increasing values of $v$ a phase shift
$\Delta \phi$  between the periodic forcing $\sim \cos(qx)$ and the 
solution  $\psi_s(x)$ occurs. This is shown in Fig.~\ref{bifv1} for three different
values of $v$ for both
the $A_1$- and the $A_3$-solution. The $A_1$-solution, which is in antiphase with
the external modulation, drags behind the forcing ($\Delta \phi > 0$). This delay increases
 with increasing values of $v$ (all other parameters fixed).
The maximum phase shift that can be achieved is about $\Delta \phi_m=\pi/2$ at a
certain velocity $v_{ex}$ above which the solution does not exist.
In contrast to this
behavior the $A_3$-solution has a negative phase shift $\Delta \phi<0$ 
with respect to the forcing as indicated by the 
lower part in Fig.~\ref{bifv1}. The maximum absolute value of the phase shift in this case is also $\pi/2$.
As indicated in Fig.~\ref{bifv2} (b) the critical phase shift $\Delta \phi_m = \pi/2$ for the $A_1$-solution 
is reached with an increasing value of the  velocity $v$ 
at lower values of the control parameter $\varepsilon$.

For a fixed $\varepsilon$ a
velocity $v_{ex}$ can be determined by the $\pi/2$-criterion where both the $A_1$- and the $A_3$-solution 
vanish to exist whereas the
$A_2$-solution still exists and propagates ahead of the external modulation with the absolute value of the phase shift always smaller than $\pi/2$. 
However, as will be shown later, this solution is always unstable.

The amplitudes for the three solutions $A_i$ ($i=1,2,3$) are plotted as a function
of the control parameter $\varepsilon$  for four different values
of the propagation velocity $v$ in the bifurcation diagram in Fig.~\ref{bifv2}. As can be seen 
in this plot, 
the amplitudes for all four velocities are nearly identical.

The shape of the solution
$\psi_s$ becomes steplike for small values of $q$, as indicated in Fig.~\ref{bif2} (a). In the
range $q\lesssim0.3$ the solution can be approximated rather well by
a series of hyperbolic tangents as shown in Appendix \ref{appA2}.
In the opposite case
$q>0.5$ one can achieve a rather good approximation of the
periodic solution by restricting the
Fourier ansatz in Eq.~(\ref{Fourieransatz}) to the leading modes as explained
in Appendix \ref{appA3}.

\section{Stability of periodic solutions in 1D and 2D}\label{stabchap}
The stationary periodic solutions $\psi_s(x)$ of Eq.~(\ref{basicequation2})
studied so far are stable only in a subrange of parameters
where they exist. In this section
we investigate the stability of periodic solutions $\psi_s(x)$ of Eq.~(\ref{basicequation2}) with respect 
to small perturbations. The resulting linear partial differential equation for the perturbation $\psi_1$
\begin{eqnarray}
\label{linstab}
\partial_t \psi_1(x,t) = \partial_x^2 \left(-\varepsilon  +3 \psi_s^2 - \partial_x^2  \right) 
\psi_1+v\partial_x\psi_1
\end{eqnarray}
contains a spatially periodic coefficient depending on $\psi_s$.
Accordingly the solutions of Eq.~(\ref{linstab}) are due to the Floquet theory 
of the form
\begin{eqnarray}\label{ansatz2}
\psi_1(x,t) = e^{\sigma t} ~e^{isx} ~ \phi_F(x)\,,
\end{eqnarray}
with the Floquet parameter $s$ and a $2\pi/q$-periodic function $\phi_F(x)$. 
This periodic function $\phi_F(x)$ can be represented by a Fourier expansion
\begin{eqnarray}\label{phiF}
 \phi_F(x)= \sum_{n=-N}^N ~ D_n ~ e^{iqxn}.
\end{eqnarray}
Taking into account Eq.~(\ref{Fourieransatz}) for $\psi_s$ the linear partial differential equation (\ref{linstab}) 
is transformed into an eigenvalue problem
\begin{eqnarray}
 \sigma D_l&=&\left[\varepsilon(ql+s)^2-(ql+s)^4+iv(ql+s)\right]D_l\nonumber \\
&&-3(ql+s)^2\sum_{k,m}C_k C_m D_{l-k-m}\,,\,\,\,\,\,\,l=-N..N,\qquad\label{ew}
\end{eqnarray}
with the coefficients $C_k$ determined by Eq.~(\ref{statlsg}).

\begin{figure}[ht]
 \centering
  \begin{minipage}[b]{7cm}
\includegraphics[width=0.85\columnwidth]{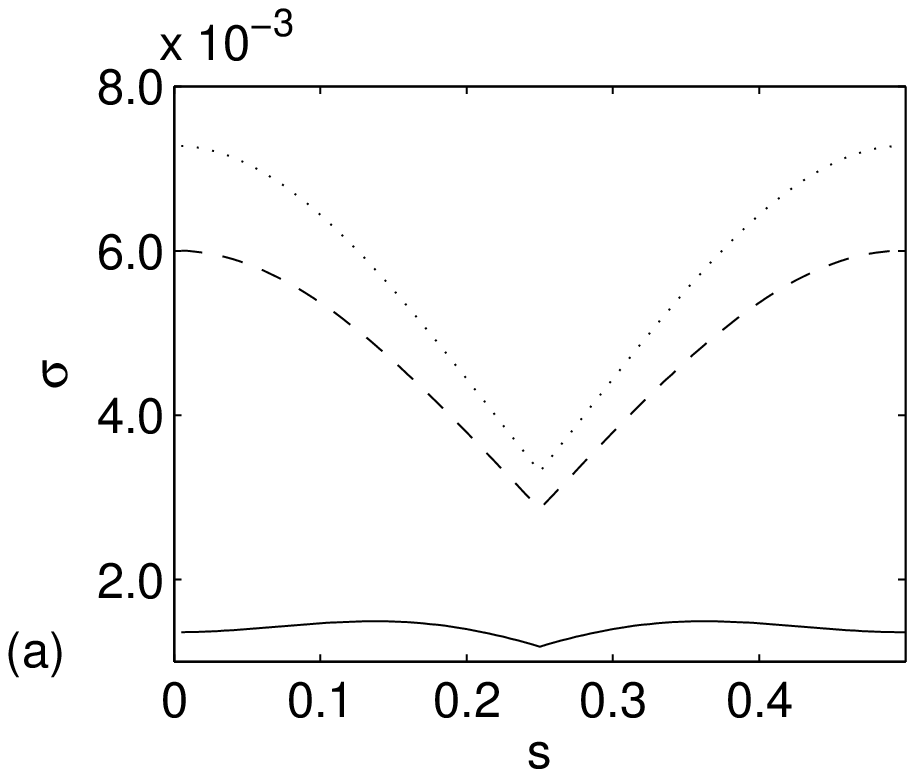}
  \end{minipage}
\begin{minipage}[b]{7cm}
\includegraphics[width=0.85\columnwidth]{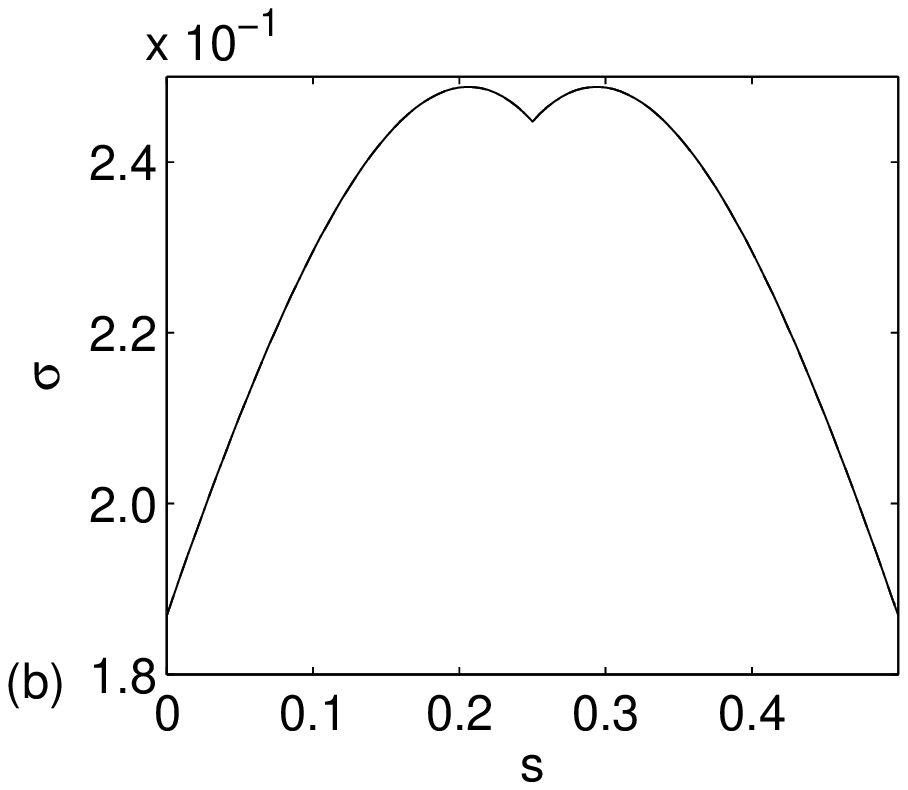}
  \end{minipage}
\begin{minipage}[b]{7cm}
\includegraphics[width=0.85\columnwidth]{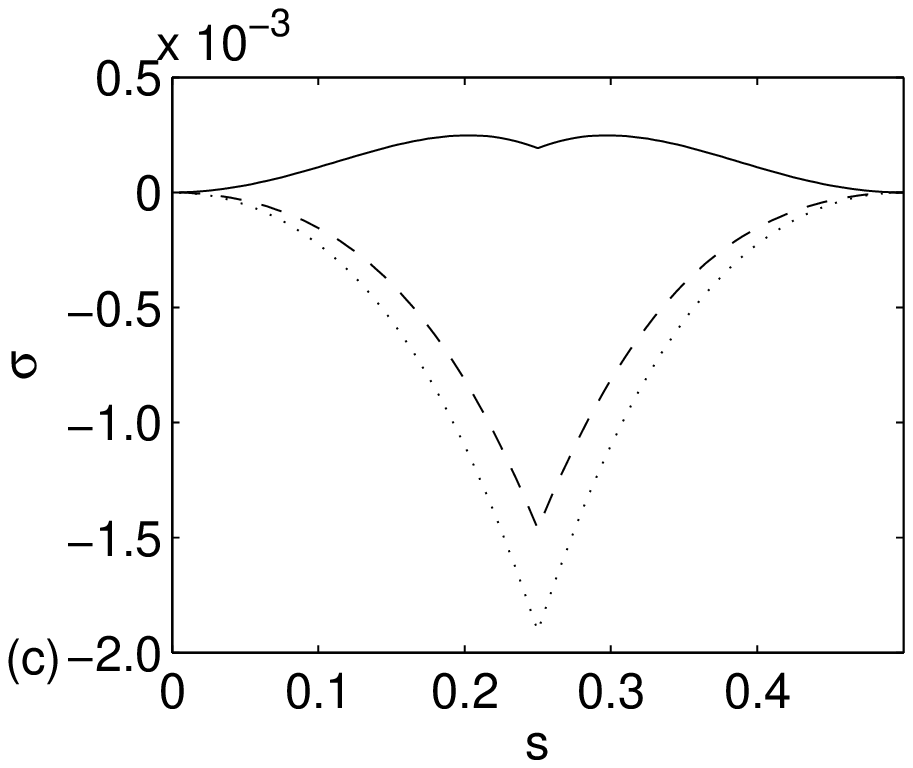}
  \end{minipage}
  \caption{The growth rate $\sigma$ of small perturbations with
respect to the stationary periodic solution is shown as a function of the
Floquet parameter $s$ for the $A_1$ (a), the $A_2$ (b) and the $A_3$-solution (c) for the
three propagation velocities $v=0$ (dotted), $v=0.008$ (dashed) and
$v=0.014$ (solid). Further parameters are 
$q=0.5$, $a=0.03$ and $\varepsilon=1$. In case of the $A_2$-solution all three lines are indistinguishable.
}
\label{stabv1}
\end{figure}

\begin{figure}[h!]
 \centering
  \includegraphics[width=0.83\columnwidth]{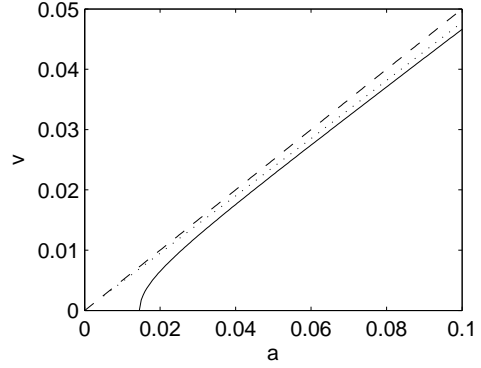}
\caption{\label{stabv2}Above the solid line the
spatially periodic solution is unstable. The dashed 
line marks the existence boundary above which the
spatially  periodic solution does
not exist due to the criterion given by Eq.~(\ref{bedingung}). The 
dotted curve marks the existence boundary obtained numerically from Eq.~(\ref{statlsg}).
The parameters are $q=0.5$ and $\varepsilon =1$.}
\end{figure}

The growth rate $\sigma$ for the solutions $A_1$, $A_2$ and $A_3$ is shown in Fig.~\ref{stabv1} as function of the Floquet
parameter $s$ for three different values of the velocity $v$ and fixed
values of the forcing amplitude $a$, the forcing wavenumber $q$ and
the control parameter $\varepsilon$.
We found that for the solutions $A_2$ and $A_3$ the growth rate $\sigma$ is positive 
for all combinations of the parameters $\varepsilon$, $q$, $a$ and $v$. Accordingly $A_2$ and $A_3$ are always unstable. 
The growth rate of the perturbation of the $A_1$-solution is negative for certain parameter ranges, meaning that
this is the only solution that can be stabilized. For fixed $\varepsilon$ and $q$ the modulation amplitude $a$
has to exceed a certain value $a_s(q)$ to stabilize the $A_1$-solution. If the traveling
velocity $v$ is smaller than a critical one $v_c(\varepsilon,a,q)$ the solution remains
stable.
The critical velocity $v_c(\varepsilon,a,q)$ is given by the solid line in Fig.~\ref{stabv2}
for a certain set of parameters and for $v>v_c(\varepsilon,a,q)$ the periodic solutions
are linearly unstable.
The onset of instability occurs for small values of the Floquet exponent $s\rightarrow0$,
i.e. belongs to a long-wave perturbation.
In Fig.~\ref{stabv2} also
the boundary of the existence range of periodic solutions $v_{ex}(\varepsilon,a,q)$ is 
shown as obtained approximately by Eq.~(\ref{bedingung}) (dashed line)
 and by a full numerical simulation (dotted line). Since the stability
boundary (solid line) always lies below $v_{ex}$ the 
periodic solution always becomes unstable before the 
existence range is reached. 
\\
Extending the analysis to two spatial dimensions the stability of a two-dimensional stripe pattern
that is periodic in the $x$-direction can be investigated
using $\psi(x,y,t)=\psi_s(x)+\psi_2(x,y,t)$ with $|\psi_2|<<|\psi_s|$. The procedure
is the same as in one dimension whereas the perturbation $\psi_2(x,y,t)$
\begin{eqnarray}
\psi_2(x,y,t) = e^{\sigma t} ~e^{i(sx+py)} ~ \phi_F(x)\,,
\end{eqnarray}
depends now on two Floquet parameters $s$ ($x$-direction) and $p$ ($y$-direction).
Linearizing Eq.~(\ref{basicequation2}) and using expansion (\ref{phiF}) one obtains the eigenvalue problem
\begin{eqnarray}
\sigma D_l&=&[\varepsilon(ql+s)^2+\varepsilon p^2-(ql+s)^4\nonumber\\
&&-2(ql+s)^2p^2-p^4+iv(ql+s)]D_l\nonumber\\
&&-3[(ql+s)^2+p^2]\sum_{k,m}C_k C_m D_{l-k-m}\,,\,\,l=-N..N.\nonumber\\[-3.5mm]
\end{eqnarray}
The linear stability analysis shows that additional transversal degrees of freedom do not influence the stability
diagram obtained for the one-dimensional case.

\section{Dynamics of the phase separation} \label{tempevol}
To study the temporal evolution of the phase separation numerical simulations of Eq.~(\ref{basicequation1}) for one and two spatial dimensions
have been performed.
We have used a central finite difference approximation of the spatial derivatives and an Euler integration of the resulting
ordinary differential equations in time.
Periodic boundary conditions have been used and we start initially with the homogeneous state $\psi=0$ with small superimposed noise.

\subsection{1D Simulations} \label{1d}
\begin{figure}[ht]
  \begin{center}
\includegraphics[width=0.74\columnwidth]{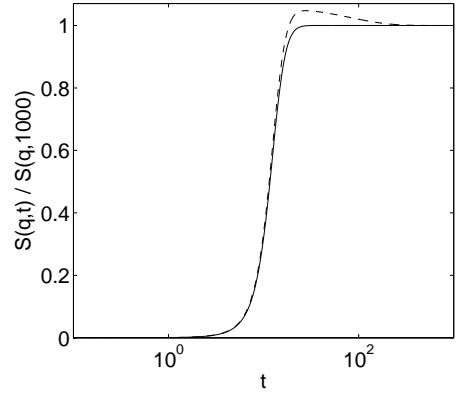}
  \end{center}
  \caption{\label{struktfakt}The dynamics of the maximum of the structure factor (relating to
its final value) for $v=0$ (solid line)
and $v=0.0266<v_c$ (dashed line). The other parameters are $\varepsilon=1$, $q=0.5$ and $a=0.1$.}
\label{temp1}
\end{figure}

In one dimension we have discretized a system of length $L=24\,\pi/q$ by $N=200$
steps with a step size $\delta x=L/N$ and a time step $\Delta t=10^{-3}$.
The noise amplitude for the initial condition $\psi=0$ was adjusted at $10^{-4}$.
The evolution of the inhomogeneous pattern has been characterized by the structure factor at the
modulation wavenumber $q$
\begin{equation}
S(q,t)=|\hat{\psi}(q,t)|^2\,,\,\hat{\psi}(q,t)=\int dx e^{iqx}\psi(x,t).
\end{equation}
In the early stages of phase separation
the amplitude of $S(q,t)$ grows faster for $v\neq 0$ and even exceeds its final value for some time 
as can be seen in Fig.~\ref{struktfakt}.
\begin{figure}[h!]
 \centering
 \begin{minipage}[b]{7 cm}
  \includegraphics[width=7.5cm,height=5cm]{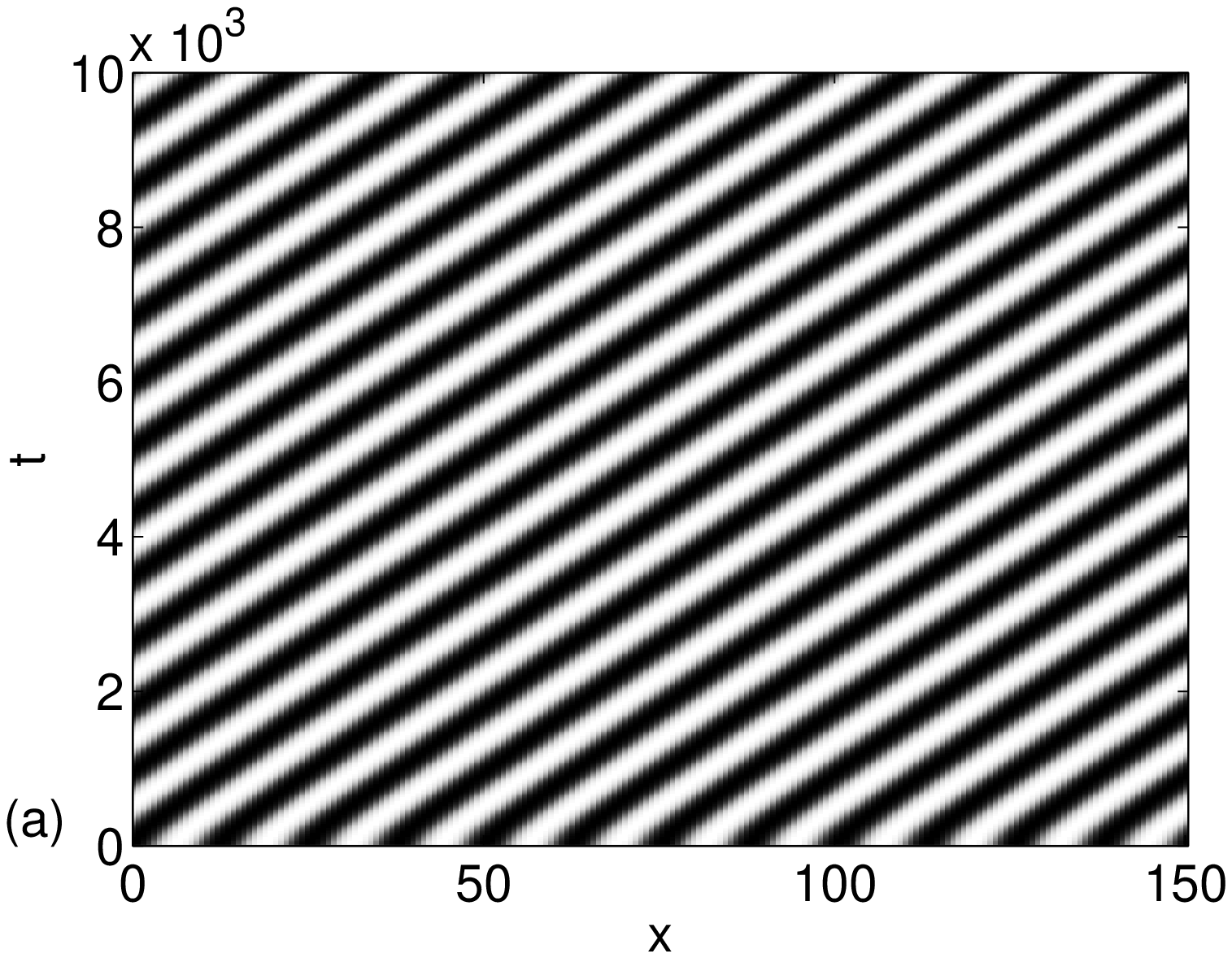}
 \end{minipage}
 \begin{minipage}[b]{7 cm}
  \includegraphics[width=7.5cm,height=5cm]{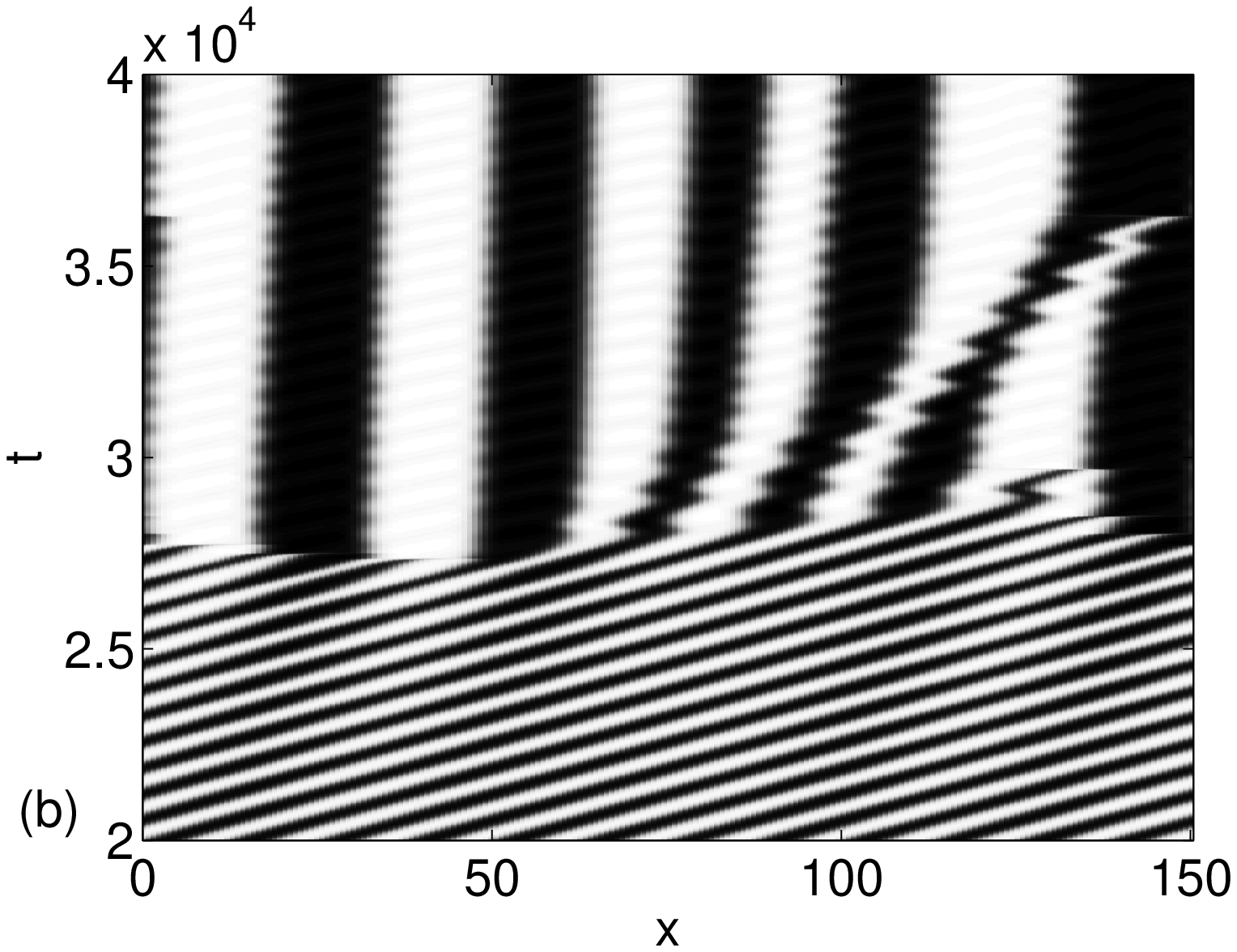}
 \end{minipage}
 \begin{minipage}[b]{7 cm}
  \includegraphics[width=7.5cm,height=5cm]{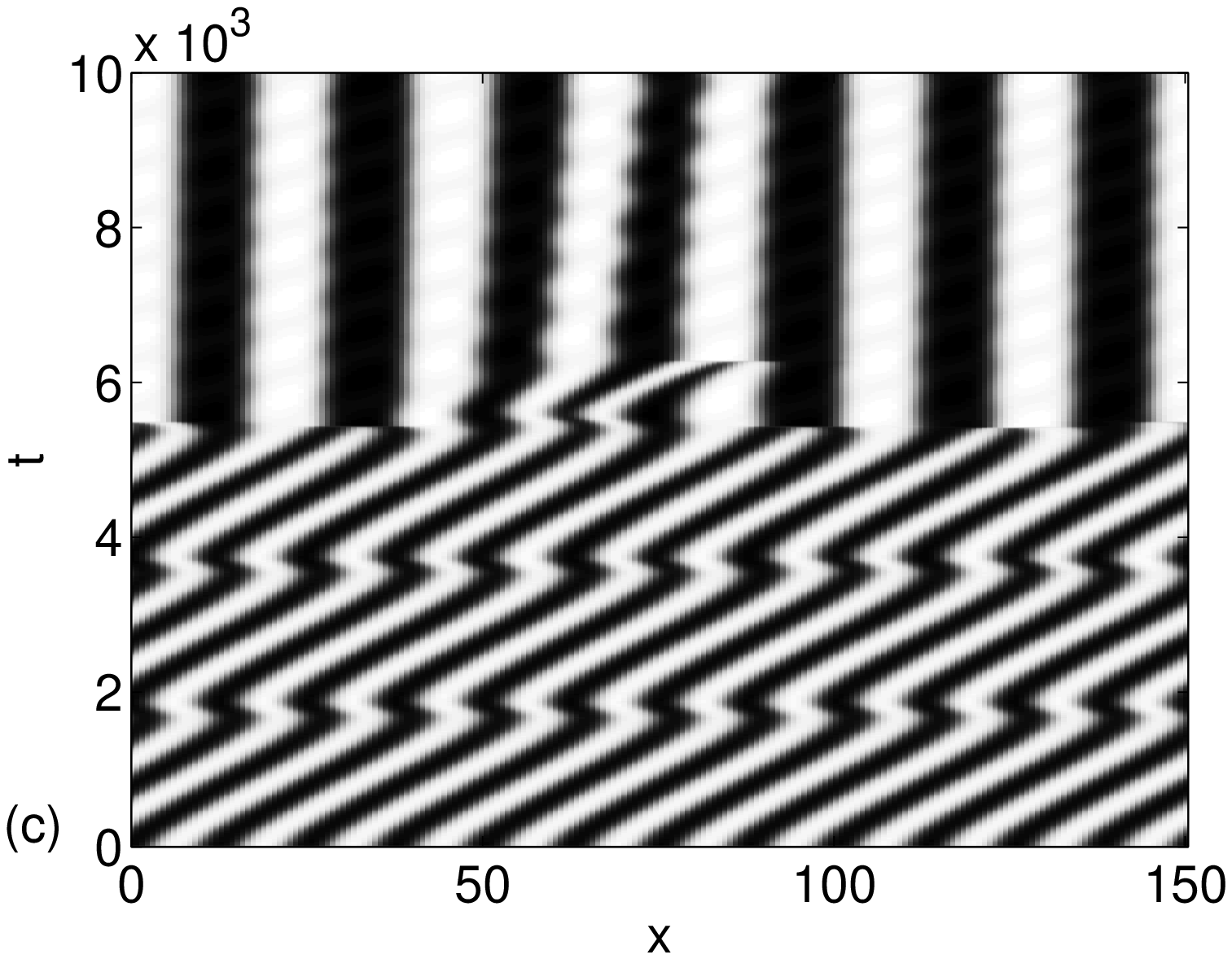}
 \end{minipage}
\begin{minipage}[b]{7 cm}
  \includegraphics[width=7.5cm,height=5cm]{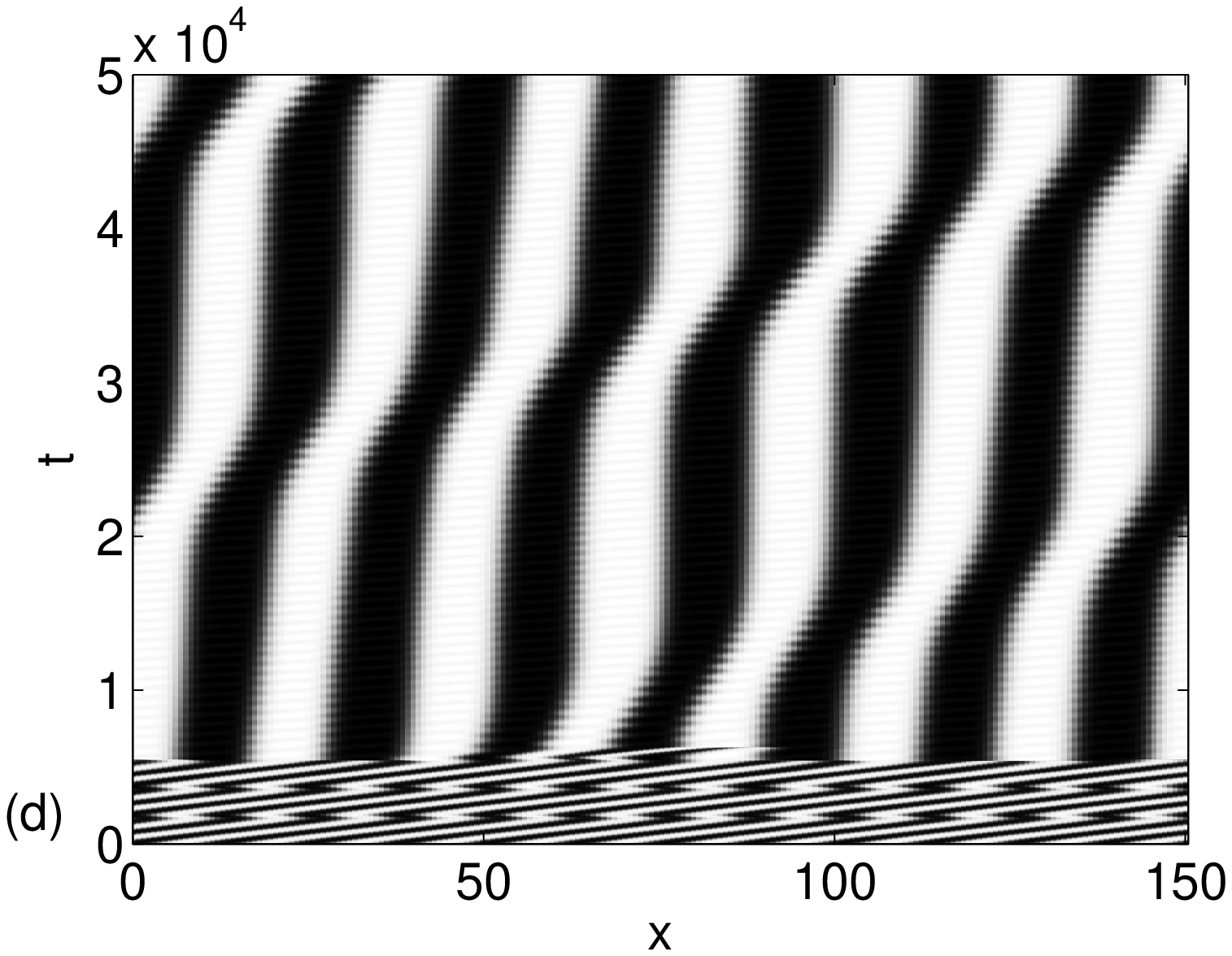}
 \end{minipage}
\caption{\label{time}Temporal evolution of phase separation is shown for $v=0.015<v_c$ (a), 
$v_c<v=0.0185<v_{ex}$ (b) and
 $v=0.02>v_{ex}$ (c,d). The other parameters are $\varepsilon=1$, $a=0.04$ and $q=0.5$.}
\end{figure}
The temporal evolution of $\psi(x,t)$
is shown in Fig.~\ref{time} for three velocity regimes. In part (a) of Fig.~\ref{time}
the velocity is sufficiently small and belongs to the range where the pattern
is locked to the propagating external modulation. In part (b) the velocity is chosen in
the range $v_c<v=0.0185<v_{ex}$ where the solution locked to the traveling external modulation still
exists, but where it is linearly unstable. In this parameter range the solution is locked
during the initial period of phase separation before coarsening takes over. In parts
(c) and (d) the temporal evolution of phase separation is shown for a velocity $v=0.02>v_{ex}$ where
the locked solution does not exist anymore. At this velocity an interesting
pinning-depinning behaviour can be observed during the initial stage of phase separation.
One still has a travelling periodic solution with the same wavenumber as the forcing but with a velocity smaller than the velocity of the forcing.
Due to the velocity mismatch the phase shift between the solution and the forcing is slowly increased
before it reaches about half of the forcing period. From that moment the periodic solution practically stops
moving (pinning) until the forcing shifts over the next half of the period.
After that the solution starts moving again (depinning) and the process repeats itself a few times.
Later on the agreement of the wavelengths of the solution and the forcing cannot be maintained anymore and the coarsening takes place.

\subsection{2D Simulations} \label{2d}
In two spatial dimensions the same system size as in 5.1 has been used in both directions,
$L_x=L_y=24\,\pi/q$ with $N_x=N_y=256$ discretization points. Again we chose
$10^{-4}$ as noise amplitude for the initial homogeneous state.\\
The numerical simulations in two dimensions are in agreement with the results of the linear
stability analysis of Sec. 4.
\begin{figure}[h!]
\centering
 \begin{minipage}[b]{4.5 cm}
\rput(0.15,3.25){(a)}
  \hspace{4mm}\includegraphics[width=3.5cm,height=3.5cm]{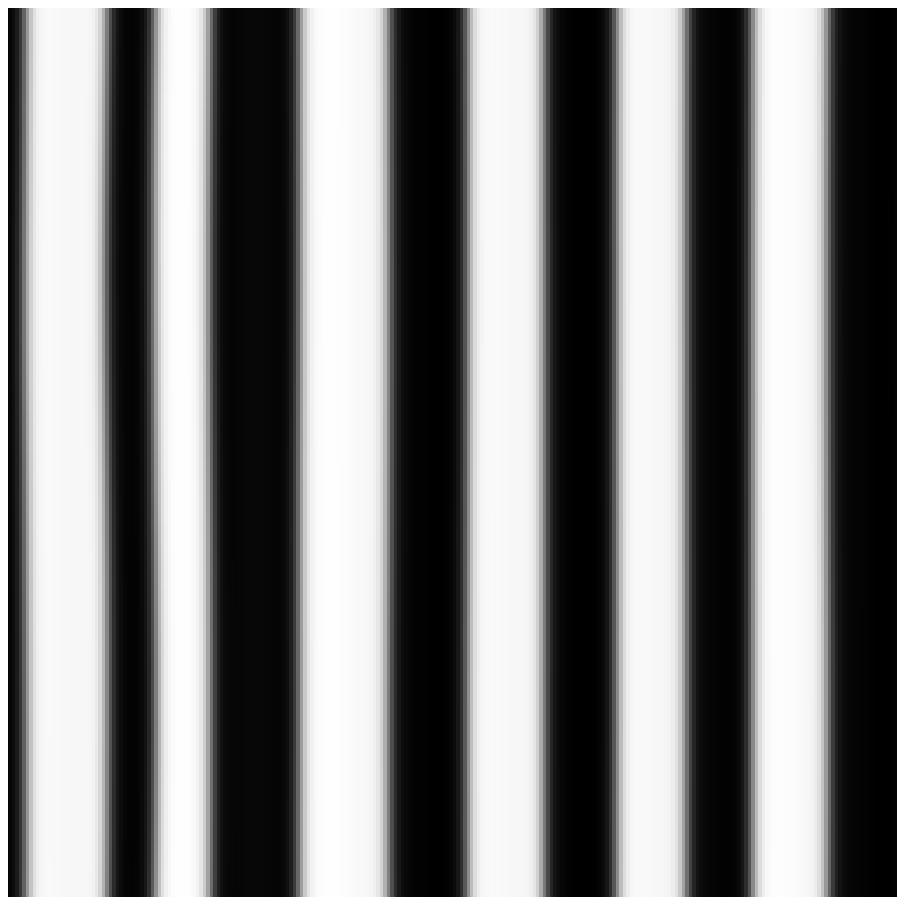}
 \end{minipage}
 \begin{minipage}[b]{4 cm}
\rput(-0.25,3.25){(b)}
 \includegraphics[width=3.5cm,height=3.5cm]{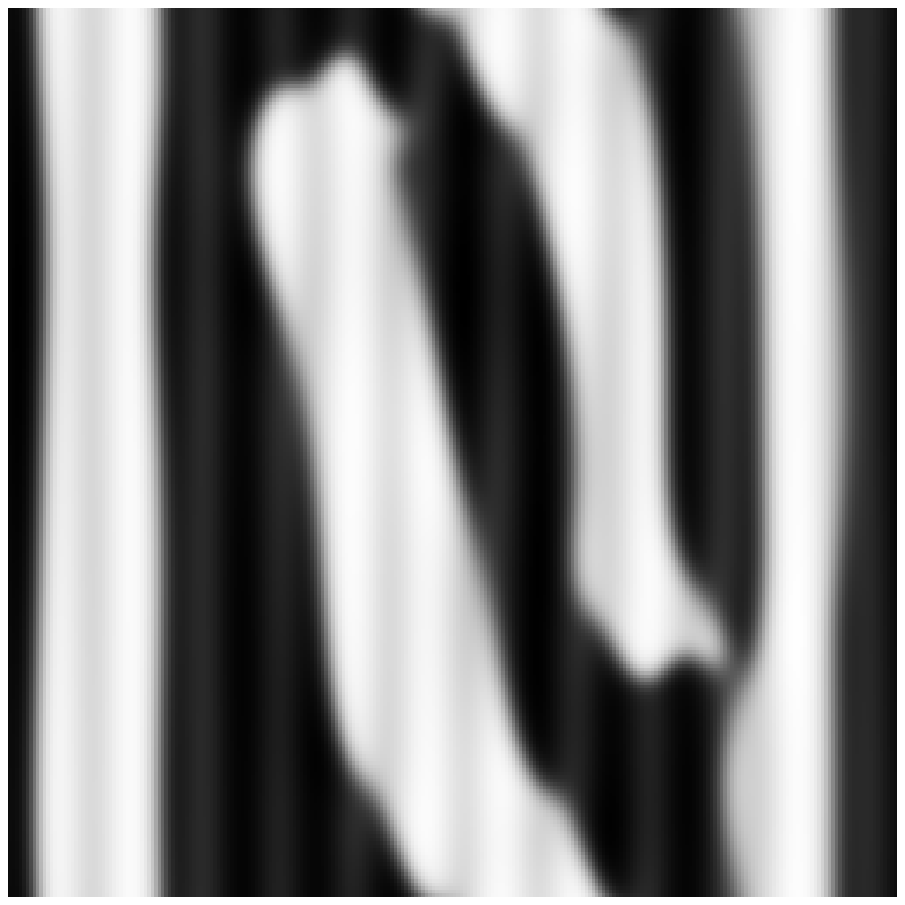}
 \end{minipage}
 \begin{minipage}[b]{4.5 cm}
 \hspace{-1.6mm}\includegraphics[width=4.5cm,height=1cm]{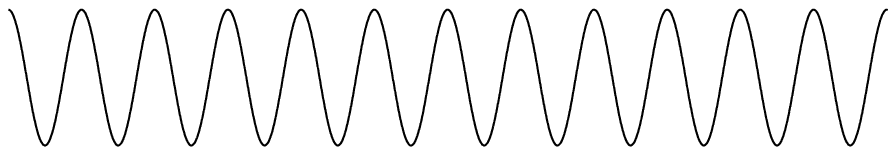}
 \end{minipage}
 \begin{minipage}[b]{4 cm}
 \hspace{-5.8mm}\includegraphics[width=4.5cm,height=1cm]{Fig9c.eps}
 \end{minipage}
 \caption{\label{2D}Snapshot of the phase separation in 2d for $q=0.5$, $v=0.0188\,(\approx1.07\cdot v_c)$, $a=0.04$, $\varepsilon=1$ at $t=5\times10^{4}$ (a) and 
 for $q=0.75$, $v=0.266\,(\approx1.32\cdot v_c)$, $a=0.3$, $\varepsilon=1$ at $t=5\times10^{3}$ (b). The curves at the bottom indicate the external forcing.}
\end{figure}\\
For $a>a_c$ and $v<v_c$ one obtains stable stripe patterns. In the unstable range
$v>v_c$ the locking of the solution gets lost and a coarsening sets in as shown in Fig.~\ref{2D} (a)
for $v\approx1.07\cdot v_c$ slightly beyond the critical velocity and in Fig.~\ref{2D} (b) for $v=0.266$ ($\approx 1.32\cdot v_c$).
For $v>v_c$ the linear stability analysis of the locked solutions in two spatial dimensions shows that transversal perturbations
being periodic in the $y$-direction exhibit also a positive growth rate $\sigma(s,p)$. However, for small values of the forcing wavenumber $q$,
the modulational wavenumber $p$ is very small and the corresponding length $2\pi/p$ may be larger than the system size in our simulations.
In this case the coarsening is quasi-one-dimensional along the $x$-direction as can be seen for instance in Fig.~\ref{2D} (a).
For larger values of $q$ also the length scale $2\pi/p_c$, with $p_c$ at the maximum of the growth rate $\sigma(s,p)$, becomes smaller
than the system length $L_y$. In this parameter range the coarsening shows a two-dimensional behaviour as shown in Fig.~\ref{2D} (b).

\section{Summary and conclusions}\label{sec: conclusions}
Effects of a spatially periodic forcing,  which travels with
velocity $v$, on  phase separation in binary mixtures 
have been investigated in terms of a generalized
Cahn-Hilliard model.  For a sufficiently large
forcing amplitude we identified three characteristic velocity 
regimes which are distinguished by the different evolving pattern and
their dynamics. For a 
vanishing velocity $v$ and the forcing amplitude $a$ larger than 
a critical value $a_c$ the phase separation can be
locked to the external modulation \cite{Krekhov:2004.1}.

For a finite velocity $v$ a condition is derived in 
Appendix \ref{appA1} which determines the parameter range 
where the spatially periodic solutions of the 
same wavelength as the external forcing  exist. 
This spatially periodic solution is locked with respect to
the external forcing in a finite parameter
range beyond
$a > a_c$ and it  is dragged by the forcing with
some phase delay. 
 This phase delay increases with the velocity $v$ up to some critical value $v_{ex}$. 
Beyond $v_{ex}$ the locked spatially periodic solution does not exist anymore
as described in Section \ref{statprop} and as shown by Fig.~\ref{bifv2}.

In the limit of small values of the modulation wavenumber $q$ 
the forced solutions resemble a periodic
array of kinks and may
be approximated by a series of hyperbolic tangents as given
by Eq.~(\ref{ansatz}). For this approximate solution we
derived in Appendix \ref{appA2} with the help of the
existence condition given in Appendix \ref{appA1} an analytical
expression for the boundary of the parameter range of 
existence 
of the locked solutions. Also in the range of larger
values of $q$ an analytical approximation of
the nonlinear solution as well as its boundary of the
existence range has been determined in Appendix \ref{appA3}.

The locked and dragged spatially periodic solutions are linearly
stable only in a subdomain of their range of existence as
described in Section \ref{stabchap}. The critical velocity $v_c(\varepsilon, a,q)$,
beyond which the locked solutions become unstable and
a depinning of the locked solutions takes place, is given as
function of $a$ in Fig.~\ref{stabv2}. For large values of $q$ 
the numerical results of the stability analysis have been
compared with results of an analytical approximation 
in Appendix \ref{appA3}, cf. Fig.~\ref{linstab2}.

In the parameter range of unstable locked solutions
one observes in the two phase region a coarsening dynamics
rather similar to the coarsening in the 
unforced case. However, during an initial transient
period the nonlinear evolution of the
solutions is different in the parameter range where
the locked solutions still exist from that where they
do not exist anymore. In the former case they are just locked during
a transient period
to the traveling forcing and in the latter case 
irregular phase jumps occur. 

In the range, where the locked solutions are either unstable or do not
exist, we observe for finite values of the traveling velocity $v$ a
faster coarsening dynamics than in the limit $v=0$.

The analysis presented here shows that travelling spatially periodic forcing
can serve as an effective tool for the control of phase separation.
The finite velocity of the forcing leads to a faster dynamics of the initial
stage of phase separation compared to the case of a stationary spatially periodic forcing.
The results should be relevant to thin polymer blends with a moderate Soret coefficient 
and a weak periodic temperature modulation created, e.g., by means of optical
grating techniques.

\section*{Acknowledgments}
We thank C. Feller, W. K\"ohler and W. Pesch for enlightening discussions.
Financial support by DFG through SFB 481 is gratefully acknowledged.

\appendix
\section{Stationary periodic solutions}
\subsection{Existence of spatially periodic solutions}
\label{appA1}
The condition of existence of stationary periodic solutions $\psi_s(x)$
of Eq.~(\ref{basicequation2}) with the wavenumber $q$ of the external forcing is derived in
this section. $\psi_s(x)$ is a solution of the following equation
\begin{equation}
 0=\partial_x^2[-\varepsilon\psi_s+\psi_s^3-\partial_x^2\psi_s+a\cos(qx)]
  +v\,\partial_x\psi_s.  \label{statgleichung}
\end{equation}
Assuming $\psi_s(x)$ being a periodic function with a vanishing mean value $\int_{0}^{2\pi/q} \psi_s(x')dx'=0$
one obtains after two integrations of Eq.~(\ref{statgleichung})
\begin{eqnarray}\nonumber
\label{help2eq}
&& \partial_x \left\{-\varepsilon\frac{1}{2}\psi_s^2+\frac{1}{4}\psi_s^4-\frac{1}{2}(\partial_x\psi_s)^2\right\}+a\cos(qx)(\partial_x\psi_s)\nonumber \\
&& \qquad +v\,(\partial_x\psi_s)\int\limits_{0}^{x}\psi_s(x')dx'=C\,(\partial_x\psi_s).
\end{eqnarray}
Using again the periodicity of $\psi_s$, an integration of (\ref{help2eq}) with respect to
the interval $[0, 2\pi/q]$
gives 
\begin{equation}
 0=\int\limits_0^{2\pi/q}dx \left[ a\cos(qx)(\partial_x\psi_s)
+v(\partial_x\psi_s) \int\limits_0^x\psi_s(x')dx' \right].\label{help3eq}
\end{equation}
After an integration by parts of (\ref{help3eq})
the constraint of existence for stationary periodic solutions $\psi_s(x)$ of Eq.~(\ref{basicequation2}) follows:
\begin{equation}
 0=aq\int\limits_0^{2\pi/q}\sin(qx)\psi_s(x)dx - v\int\limits_0^{2\pi/q}\psi_s^2(x)dx. \label{existenz}
\end{equation}

\subsection{Analytical approximation in the limit of small $q$}
\label{appA2}
For small values of the
wavenumber $q$ the solutions for the $A_1$-branch  become steplike periodic functions as indicated in Fig.~\ref{bif2} (a).
Kink-type periodic functions
may be approximated by a series of hyperbolic tangents \cite{Langer:71.1}
\begin{equation}
 \psi_s(x) = \sqrt{\varepsilon}\cdot\sum_{n=0}^{N} (-1)^n \tanh\left[M(x) \right]\label{ansatz}
\end{equation}
where
\begin{eqnarray}
M(x)= \sqrt{\frac{\varepsilon}{2}}~\left[x-(n+\frac{1}{2})\frac{\pi}{q}+\gamma\right]
\end{eqnarray}
describing a periodic array of $N$ single interfaces with a spacing $\pi/q$ (size of 
the system $L=N\cdot\pi/q$). $\gamma$ depends
on $q, ~ \varepsilon, ~ a$ and $v$.
Inserting the ansatz (\ref{ansatz}) into Eq.~(\ref{existenz}) one obtains
\begin{eqnarray}\nonumber
&& \int\limits_0^{2\pi/q} \sin(qx)\cdot \sqrt{\varepsilon}\sum_{n=0}^{N} (-1)^n \tanh\left[ M(x)\right] dx \\ &&\quad=\frac{v}{aq}\int\limits_0^{2\pi/q}\Bigg[\sqrt{\varepsilon}\sum_{n=0}^{N} (-1)^n \tanh\left[M(x)\right]\Bigg]^2 dx
\end{eqnarray}
for the determination of $\gamma$, which may be reformulated in the following form
\begin{eqnarray}\nonumber
&& \int\limits_0^{2\pi/q} \sin(qx)\cdot \sum_{n=0}^{N} (-1)^n \tanh\left[M(x)\right] dx\\\label{umform} &&\quad\quad=\frac{v\sqrt{\varepsilon}}{aq}\Bigg\lbrace\frac{2\pi}{q}-\int\limits_0^{2\pi/q}\sum_{n=0}^{N}  \mbox{sech}^2\left[M(x)\right]dx\Bigg\rbrace.\qquad
\end{eqnarray}
For $q\ll1$ the first integral in this equation may be approximated by the expression
\begin{eqnarray}\nonumber
 \int\limits_0^{2\pi/q} \sin(qx)\cdot  \sum_{n=0}^{N} (-1)^n \tanh\left[M(x)\right] dx \approx \frac{4}{q}\sin(\gamma q)\\
\end{eqnarray}
and the second integral by
\begin{eqnarray}
 \int\limits_0^{2\pi/q}\sum_{n=0}^{N}  \mbox{sech}^2\left[M(x)\right] dx\approx 4\sqrt{\frac{2}{\varepsilon}}\,.
\end{eqnarray}
Both integrals together with  Eq.~(\ref{umform}) 
give an explicit expression for the constant $\gamma$
\begin{eqnarray}
 \gamma=\frac{1}{q}\mbox{arcsin}\left[\frac{v\sqrt{\varepsilon}}{a}\left(\frac{\pi}{2q}-\sqrt{\frac{2}{\varepsilon}}\right)\right] \label{gam2}
\end{eqnarray}
in terms of the parameters of the external forcing. 
With $\gamma$ as determined above
the representation in Eq.~(\ref{ansatz}) is a good approximation
of the solution as it deviates less than 10\% from the full numerical solution for $q\lesssim0.3$.
The deviation decreases with decreasing $q$.

In the range of the coexistence of the three branches $A_1$, $A_2$ and $A_3$ 
the shift $\gamma$ between the forcing and the
spatially periodic solution increases with an increasing velocity $v$ up to the
numerically determined existence limit 
of the $A_1$-solution at approximately $\gamma q=\Delta \phi_m=\pi/2$, as described
in Sec.~\ref{statprop}.  With this condition $\gamma q=\pi/2$ 
we may also calculate 
the velocity
\begin{equation}
v_{ex} = \frac{2aq}{ \pi\sqrt{\varepsilon} -2q\sqrt{2}} 
\end{equation}
below which the periodic kink-type solution exists. This analytical
expression for $v_{ex}$ is in a good agreement with the numerically determined
existence boundary.

\subsection{Analytical approximation in the limit of large $q$}
\label{appA3}
For large $q$ the nonlinear solution can be
described by a leading order Fourier expansion: $\psi_s \propto \cos(qx)$.
However, if the forcing is propagating with a velocity $v$ a phase
shift between the forcing and the periodic solution $\psi_s(x)$ 
occurs and this phase shift can be taken into account
by the following ansatz:
\begin{equation}
 \psi_s(x)= C_1\cos(qx)+C_2\,v\sin(qx)\label{zweimoden}.
\end{equation}
This corresponds to a restriction of the series given by
Eq.~(\ref{Fourieransatz}) to $N=1$. Inserting the ansatz (\ref{zweimoden}) 
in Eq.~(\ref{statgleichung}) and collecting the linear independent 
contributions proportional to $\sin(qx)$ and $\cos(qx)$ one obtains
the two coupled equations for the coefficients $C_1$ and $C_2$
\begin{eqnarray}\label{c1}
 \left( \varepsilon - q^2 -\frac{3}{4}\,C_1^2 -\frac{3v^2}{4}\,C_2^2\right) C_1\,+\frac{v^2}{q}\,C_2-a &=&0,\\[1mm]\label{c2}
 \left( \varepsilon-q^2 -\frac{3v^2}{4}\,C_2^2-\frac{3}{4}\,C_1^2\right)C_2-\frac{1}{q}\,C_1 &=&0.
\end{eqnarray}
In the range $q>0.5$ a numerical solution of these two coupled equations provides solutions which deviate less than 30\% from the full
numerical solution. 

\begin{figure}[ht]
\begin{center}
\includegraphics[width=0.9\columnwidth]{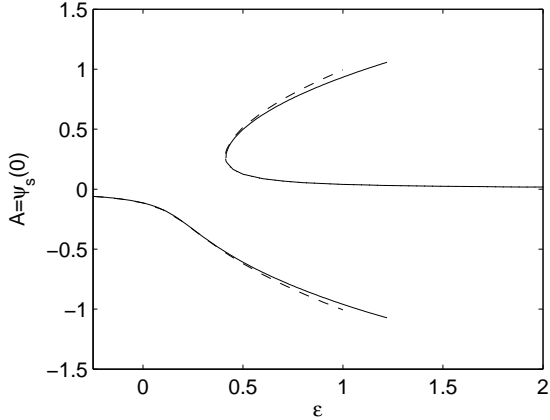}
\caption{\label{ana_bifi}The bifurcation diagram obtained from the full numerical solution (solid) and the approximation (\ref{zweimoden})
(dashed) for the parameters $q=0.5$, $a=0.03$ and $v=0.01424$.}
\end{center}
\end{figure}

The deviation decreases further with increasing $q$. A comparison between the solution (\ref{zweimoden})
and the full numerical solution is shown in Fig.~\ref{ana_bifi} for $q=0.5$. The deviations
increase at larger values of $\varepsilon$ and close to
the boarder where the stationary solutions vanish to exist. 
The existence range of the $A_1$-solution again may be determined by assuming
a maximum phase shift of $\pi/2$ between the external forcing and the periodic
solution. In this case the $\cos$-contribution in Eq.~(\ref{zweimoden}) vanishes,
that means one should have $C_1=0$. Then the velocity $v_{ex}$ can be calculated from the existence of the solution $C_2\neq0$ of Eqs.~(\ref{c1}) and~(\ref{c2}):
\begin{equation}
 v_{ex}=\frac{\sqrt{3}}{2}\,\frac{aq}{\sqrt{\varepsilon-q^2}}\,.\label{bedingung}
\end{equation}
The comparison with the numerical boundary is shown in Fig.~\ref{stabv2}.

\begin{figure}[ht]
\begin{center}
\includegraphics[width=0.9\columnwidth]{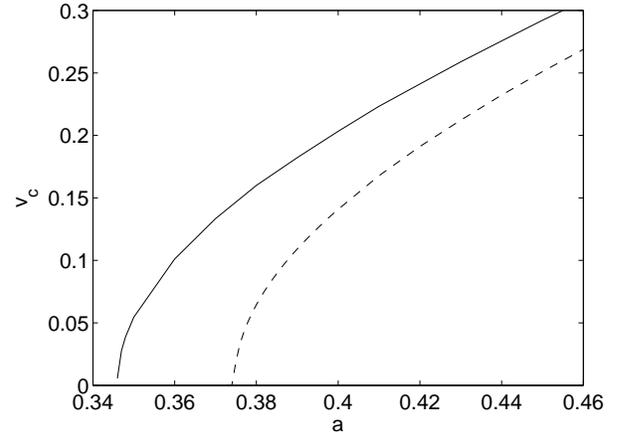}
\caption{\label{linstab2}Critical velocity from the numerical stability analysis (solid) and the
analytical approximation (\ref{c1}), (\ref{c2}) and (\ref{anastab}) (dashed) for $q=0.9$ and $\varepsilon=1$.}
\end{center}
\end{figure}

In the limit of large $q$ it is also possible to obtain an analytical approximation
for the linear stability of the locked periodic solutions.
Therefore we use Eq.~(\ref{ansatz2}) as an ansatz for the perturbation $\psi_1(x,t)$. Now we
approximate $\phi_F(x)$ by the three dominant Fourier-modes
\begin{equation}
 \phi_F(x)=D_{-2}e^{-2iqx}+D_0+D_2e^{2iqx}. \label{fansatz}
\end{equation}
Inserting (\ref{zweimoden}) and (\ref{fansatz}) into Eq.~(\ref{linstab}) we gain an eigenvalue problem.
Then we expand the growth rate $\sigma$ in terms of the Floquet parameter $s$ and use the
curvature for $s\rightarrow0$ as a criterion for the stability boundary, this leads to
\begin{eqnarray}\nonumber
&&0=27q^2C_2^6v^6-180q^2C_1^2C_2^2v^2-256q^6+128q^4-4v^2\\\nonumber
&&\qquad -16q^2-384q^4C_1^2+72q^2C_1^2-90q^2C_1^4-384q^4C_2^2v^2\\\nonumber
&&\qquad -90q^2C_2^4v^4+72q^2C_2^2v^2+384q^6C_2^2v^2+216q^4C_2^4v^4\\\nonumber
&&\qquad +384q^6C_1^2+81q^2C_2^4v^4C_1^2+81q^2C_1^4C_2^2v^2\\\nonumber
&&\qquad +432q^4C_2^2v^2C_1^2+6C_2^2v^4+6C_1^2v^2+216q^4C_1^4\\
&&\qquad +27q^2C_1^6\label{anastab}
\end{eqnarray}
as a condition for linear stability. Together with Eqs.~(\ref{c1}), (\ref{c2}) this gives a system 
of three coupled equations that can be solved numerically. In Fig.~\ref{linstab2} this 
approximation is compared with the results of the full numerical stability analysis.

\end{document}